# Design of an Automated Intrusion Detection System incorporating an Alarm

Awodele Oludele, Ogunnusi Ayodele, Omole Oladele, Seton Olurotimi
Computer Science and Mathematics Department,
Babcock University Ilishan-Remo,
Ogun state, Nigeria

**Abstract-**Security and safety are two intertwined terms. It is a common belief that when a place or system is secure, it is safe. This paper shows a means of integrating three devices for physical intrusion detection. This paper thus suggests a means of increasing the level of security in an enclosed area with the use three of the four security layers necessary for optimum security. This paper intends to show that a system with more than one security device in place tends to prevent unauthorized access. This paper would be illustrating the implementation of this in an enclosed area whose security level must be kept on the high at all times.

## 1. INTRODUCTION

Security has been defined, by [1], as safety from harm. It is a term that has different dimensions in psychology, public safety, defense and military matters and information access. Safety, also by [1], is defined as protection from action from without or subversion from within. Security and safety are always intertwined and it is impossible to design a security system without taking into account the safety of the object or person into consideration. Security in embedded systems is usually an afterthought.

Considering the current global security environment, the importance of good physical security is difficult to ignore. Physical security services are becoming a private rather than public service i.e. individuals and organizations tend to hire private security firms and install security equipment and use the police as the back. According to the Bureau of Labor Statistics (2004), private security officers outnumber police officers by more than 2 to 1 in the United States of America. Recent reports suggest that this trend holds true for both daily security operations responding to terrorism [5], natural disasters[4]. Physical security has seen less attention and it is primarily an applied field, it has no dedicated line of research. Instead, it is scattered through fields like engineering, computer science, chemistry and physics as well as social sciences such as criminology, sociology and psychology [2]. Providing security relies on two main elements:

- Equipment or technology and
- People

Without any statistical proof, it is the belief of [2] that major funding goes more into equipment security research than people security research. Others might disagree with this point of view for example, [3] stated that General Eugene Habiger (retired), former commander of US strategic nuclear forces and advisor to the US department of Energy was quoted saying that "good security is 20% equipment and 80% people". Human security, according to [3], has the tendency to want to cover more ground than actually securing the ground. Human security has several reoccurring problems which include:
- Disgruntled employees
- Professionalism
- Job characteristics
- Security structure and climate for security

These reoccurring problems tend to disturb or halt the work to be done or carried out.

The objectives of this paper are given as follows:
- To find and report as accurately as possible the previous security systems that are in place in the world today.
- To design and construct a metal detector, digital access system and a pressure sensitive security mat.
- To successfully harmonise the operations of these three devices.
- To outline and accurately state the mode of operation and conditions for optimal performance.
- To successfully implement this security system.

The scope of this paper is limited to rooms with only one entrance and exit. Such rooms include stores and vaults. Such rooms ensure that all individuals accessing the room can only access the room and leave the room via one door. This makes keeping track of access information very convenient for the user. In computer security field, an intrusion is defined as an active sequence of related events that deliberately try to cause harm, such as rendering system unusable, accessing unauthorized information or manipulating such information. It refers to both successful and unsuccessful attempts [6]. Thus this paper is justified.

## 1.1. PROBLEM STATEMENT

Security is an important aspect of all our lives. A secure environment tends to give individuals rest of mind. However, with a global recession occurring, people have resorted to less honorable means to make ends meet. Theft rates (in a typical Nigerian University) have increased causing disharmony in the areas where such events occur. This sometimes leads to unpleasant occurrences such as fights and sometimes injuries. These injuries are either sustained when the culprit is caught and a mob action is decided upon. The aim of the proposed design is to reduce such occurrences by alerting the appropriate quarters while an unauthorized entrance into a room occurs. The relevance of this paper cannot be overemphasized.

## 1.2. PREVIOUS SECURITY SYSTEMS AND THEIR LIMITATIONS

**1.2.1 Electrical Locks** Electric locks come in many forms. The most basic is a **Magnetic Lock** (commonly called a mag lock). A large electro-magnet is mounted on the door frame and a corresponding armature is mounted on the door. When the magnet is powered and the door is closed, the armature is held fast to the magnet. Mag locks are simple to install and are very attack resistant. But mag locks are also problematic. Improperly installed or maintained mag locks have fallen on people. Also there is no mechanical free egress. In other words, one must unlock the mag lock to both enter and leave. This has caused fire marshals to impose strict codes on the use of mag locks and the access control practice in general. Other problems include a lag time in releasing as the collapsing magnetic field is not instantaneous. This lag time can cause a user to walk into the door. Finally, mag locks by design fail unlocked, that is if power is removed they unlock. This could be a problem where security is a prime concern.

**Electric Strikes** replace a standard strike mounted on the door frame and receive the latch and latch bolt. Electric strikes can be simple to install when they are designed for drop-in replacement of a standard strike. But some electric strikes require that the door frame be heavily modified. Electric strikes allow mechanical free egress: As a user leaves, he operates the lockset in the door, not the electric strike in the door frame. Electric strikes can also be either fail unlocked, as a mag lock, or the more secure fail locked. Electric strikes are easier to attack than a mag lock. It is simple to lever the door open at the strike. Often the there is an increased gap between the strike and the door latch. Latch guards are often used to cover this gap

**Electric Mortise** and **Cylindrical Locks** are drop in replacements for the door mounted mechanical locks. A hole must be drilled in the door for electric power wires. Also a power transfer hinge is used to get the power from the door frame to the door. Electric mortise and cylindrical locks allow mechanical free egress. Electric mortise and cylindrical locks can be either fail unlocked or fail locked.

**Electrified Exit Hardware**, sometimes called panic hardware or crash bars, are used in fire exit applications. The idea is that one simply pushes against the bar to open it, making it the easiest of mechanically free exit methods. Electrified exit hardware can be either fail unlocked or fail locked. A drawback of electrified exit hardware is their complexity which requires skill to install and maintenance to assure proper function.

**Motor Operated Locks** are used throughout Europe. A European motor operated lock has two modes, day mode where only the latch is electrically operated, and night mode where the more secure deadbolt is electrically operated (www.wikipedia.com).

### 1.2.2. User Authentication Systems

When implemented with a digital access system, one of the following access systems or digital authentications systems can be with an electric lock. These however are only a few of the numerous authentication devices available.

**Numerical Codes, Passwords and Passphrases:** Perhaps the most prevalent form of electronic lock is that using a numerical code for authentication; the correct code must be entered in order for the lock to deactivate. Such locks typically provide a keypad, and some feature an audible response to each press. Combination lengths are usually between 4 and 6 digits long.

A variation on this design involves the user entering the correct password or passphrase. A major hindrance however is the fact that users are capable of forgetting their codes. Forgetfulness is especially common in older people and this system will not be convenient for them. These codes are, in some cases, easy to crack.

**Security Tokens:** Another means of authenticating users is to require them to scan or "swipe" a security token such as a smart card or similar, or inter act a token with the lock. For example, some locks can access stored credentials on a personal digital assistant using infrared data transfer methods. However, just as in the case of an ATM card, the magnetic tape tends to wear off with time either resulting to time wasting in accessing a room or the inability of the user to access the room at all.

**Biometrics:** As biometrics become more and more prominent as a recognized means of positive identification, their use in security systems increases. Some new electronic locks take advantage of technologies such as fingerprint scanning, retinal scanning and iris scanning, and voiceprint identification to authenticate users. This is a very secure way of identifying a person's identity but it is limited by the occurrence of an accident or disfiguration to the part of the body used for identification.

**1.3.1. Problems with the Existing System**

This paper is written using a typical University in Nigeria as a case study. The focus of this paper is to create a security system to be implemented in any room in the institutions. Babcock University for instance, bases its physical security on only the first and second layers of physical security i.e. **Crime Prevention through Environmental Design** and **mechanical** layer which include gates, doors, and locks. This makes the system effective to some extent but largely ineffective and vulnerable to attacks and manipulation both from within and without the system. The system bases its physical security on two main features namely:
- People (Security Men and Women) and
- Structures (Walls and locks)

These two features are easy to bypass and this is evident by the high rates of theft and manipulation of security men by students to do their biddings. This paper seeks to reduce the theft rates by increasing the security of such areas by making them "safe rooms". The problems with the current security structure of rooms in Babcock University include:

**Ease of Access through Breaking of Locks**

All doors in our immediate environment are based on the cylindrical lock mechanism. This lock mechanism is very common and this fact makes it easy to breakdown by either breaking the lock or duplicating the keys. This makes the system rather unsafe for all users

**Lack of Intrusion Detection Alerts**

All intrusion alerts are based dependent on discovery by individuals i.e. either security personnel's or students. This delay gives the culprit enough time to dispose of whatever has been stolen and more than enough time to cover his tracks. This leads to a string of an ever increasing number of unsolved cases of theft. A proper intrusion detection system alerts the responsible quarters once an abnormality is discovered in the system.

**Inefficient Monitoring Method**

Monitoring one's belongings are left to the vigilance on the path of the security officials and the owners of such goods. This can prove to be ineffective considering the fact that as human beings, we tend to get bored performing monotonous tasks. This leads to the search of more exciting tasks no matter how irrelevant they might be at such times. There is also the need to takes occasional breaks to refresh one's self. A very observant thief will be able to use such minute details to his advantage.

When surveillance is continuous with no visible break, it tends to deter the less desperate thieves and thereby reducing the theft rate. An example of a continuous surveillance system is the use of closed-circuit television (CCTV).

## 2. LITERATURE REVIEW
### 2.1. EXISTING SECURITY SYSTEMS

Security is the condition of being protected against danger, loss, and criminals. In the general sense, security is a concept similar to safety(www.wikipedia.com). The slight difference between the two is an added importance on being protected outside threats or dangers. Individuals or actions that go or act against the general rules of protection are responsible for the breach of security. Security can also be seen as:

- A condition that results from the establishment and maintenance of protective measures that ensures a state of inviolability from hostile acts or influences.
- With respect to classified matter, the condition that prevents unauthorized persons from having access to official information that is safeguarded in the interests of national security.
- Measures taken by a military unit, an activity or installation to protect itself against all acts designed to, or which may, impair its effectiveness.

Security has to be compared and contrasted with other related concepts: Safety, continuity, reliability. The key difference between security and reliability is that security must take into account the actions of people attempting to cause destruction (www.wikipedia.com).There is an immense literature on the analysis and categorization of security. It is common knowledge as with all systems that the "weakest link in the chain" is the most important. This can also be incorporated in security systems. The situation is asymmetric since the *defender* must cover all points of attack while the attacker need only identify a single weak point upon which to concentrate.

### 2.2. Real Security versus Perceived Security

It is often true that what individuals perceive or believe to be real security is quite the opposite is actually a mirage of the real secure system. People tend to believe air transport is more dangerous than road travel but it is quite the opposite. Sometimes the equipment or tools are mistaken for the effect. This is often demonstrated by people who have much antivirus softwares on their systems. These softwares are incompatible and tend to create more problems than solve problems. The user however believes that with more antivirus softwares the better protected the system. This leads the user to believe he has a protected system while in reality, the system is exposed. This is a typical example of real security versus perceived security. **Security Theatre** is a phenomenon where ineffective security measures are introduced and these measures sometimes increase the real security only a little or sometimes they actually decrease the real security (www.wikipedia.com).However, perceived security can actually give a form of actual security. The perception of security acts as some form of deterrent for malicious attacks or unplanned attacks. For example, if a warning sign is shown that "Trespassers will be shot on Sight", this will deter people from trespassing even if there are no armed guards on duty. Also, often when there *is* actual security present in an area, such as video surveillance, an alarm system in a home, or an anti-theft system in a car such as a LoJack, signs advertising this security will increase its effectiveness, protecting the value of the secured vehicle or area itself. Since some intruders will decide not to attempt to break into such areas or vehicles, there can actually be less damage to windows in addition to protection of valuable objects inside. Without such advertisement, a car-thief might, for example, approach a car, break the window, and then flee in response to an alarm being triggered. Either way, perhaps the car itself, the objects inside or both aren't stolen, but with perceived security even the windows of the car have a lower chance of being damaged, increasing the financial security of its owner(s). Perceived security, however, does not guarantee the security of an area. It is important, however, for signs advertising security not to give clues as to how to subvert or manipulate that security system, for example, a burglar planning to rob a home reading the name of the manufacturer from the advertisement.

### 2.3. Concepts in Security

Certain concepts recur throughout different fields of security (www.wikipedia.com).

- Assurance - assurance is the level of guarantee that a security system will behave as expected
- Countermeasure - a countermeasure is a way to stop a threat from triggering a risk event
- Defense in depth - never rely on one single security measure alone.
- Exploit - a vulnerability that has been triggered by a threat - a risk of 1.0 (100%)
- Risk - a risk is a possible event which could cause a loss
- Threat - a threat is a method of triggering a risk event that is dangerous
- Vulnerability - a weakness in a target that can potentially be exploited by a threat

### 2.4. Attributes of a Good Security System
A good security system must have the following attributes:

- Sensitivity: the system must be sensitive enough to detect threats or changes in the environment
- Reliability: the system must be dependable i.e. it must work in the environment its placed in
- Durability: it must be "rugged" i.e. work efficiently for a long time or a reasonable period.
- Ease of deployment: it must be easy to transport and set up.

## 3. METHODOLOGY OF THIS PROPOSED SYSTEM
According to [6], "An intrusion detection system itself can be defined as the tools, methods and resources to help identify, assess and report unauthorized or unapproved network activity".
A security system is highly improved with an increased number of measures and countermeasures put in place to avoid or detect an intrusion. This involves the use of several security realms monitored by a human or a machine. This research work however is only interested in the physical security of a room. A good security system consists of four layers which are:
- Environmental design
- Mechanical and electronic access control
- Intrusion detection
- Video monitoring

In the area of study, only the first two layers are implemented making the rooms insecure and also very open to attacks from both within and without. It is the intention of this paper to propose a design for an automated intrusion detection system which will alert the human security of an ongoing attack via the means of an alarm system. The **environmental design** refers to the physical structures and personnel's put in place to monitor and handle physical threats to an area. Such include walls, security officials and security animals. This paper proposes the use of a **metal detector**. The **mechanical and electronic access control** refers to the mechanical infrastructure put in place to prevent or disturb attacks. These include doors and locks. This paper proposes to replace these with **electronic locks with identity validation systems**. **Intrusion detection** deals with alarm systems put into place or alarm triggers. This paper proposes to complement the alarm with a **pressure sensitive security mat**. This will act as a trigger for the alarm system in case the first two stages are bypassed. The final layer is called the **video monitoring**. This is a recorded **video surveillance system** which could range from a camcorder with a memory device to a hidden Closed Circuit Television (CCTV). This provides a means of identifying the culprit in case the intruder is able to escape before a response team arrives. It is important to note that these layers should be able to function independently and also work as a unit. A flow chart showing the proposed sequence of events is shown in **figure 1**:

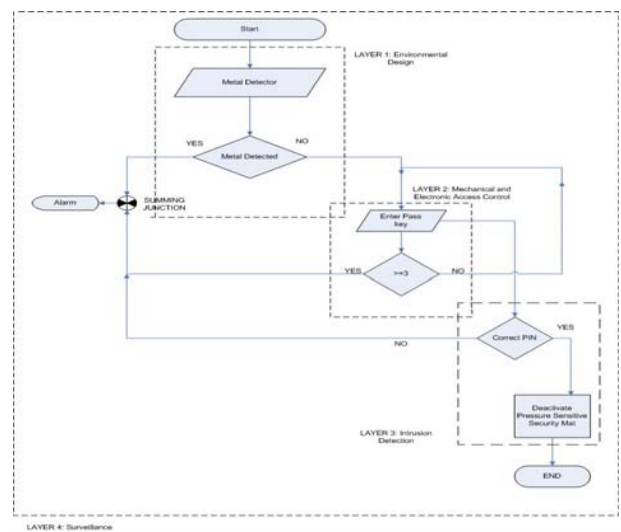

**Figure 1:** *A Flowchart Showing the Inter relationship between the Four Security Layers and the Mode of Operation of the Proposed System*

The flowchart in **figure 1** identifies four distinct layers which are represented by the devices described in this paper. This paper however is concerned with only three layers described as follows:

**3.1.0. Metal Detector (The Environmental Layer):** Metal detector technology is a huge part of our lives these days. Metal detectors are devices that use electromagnetic fields to detect and signal the presence of metallic or ferromagnetic objects. Metal detectors vary in their effective operating ranges and the amounts and types of metals necessary to generate a signal (www.howstuffwork.com). They are used in airports, schools, court houses, train stations, night clubs, special events and prisons to help ensure that no one is bringing a metallic weapon unto the premises. They are generally divided into two: **hand-held type** and the **walk through models**. The hand held type can be used alone or in conjunction with the walk through model. This can be thought of as a 'double precaution'. When a person walks through both metal detectors and the security officials are still in doubt, the person can be checked with a physical pat down. Walk through metal detectors with digital technology provide enhanced target detection coverage (www.EzineArticles.com). Multi zone walk through metal detectors are used in high security areas and feature full target coverage on the right, center and left side of the body from head to toe. Most metal detectors usually have audible and visual alarms which signal when a target has been detected.

### 3.1.1. ANATOMY OF A METAL DETECTOR

A typical metal detector is light weight and consists of just a few parts

- *Stabilizer* **(optional):** It is used to keep the unit steady as you sweep it back and forth.

- *Control Box***:** It contains the circuitry, controls, speaker, batteries and the microprocessor.
- *Shaft***:** This connects the control box and the coil, often adjustable so you can set it as a comfortable level for your height.
- *Search Coil***:** This is the part that actually senses the metal, also known as the "search head"/ "loop" / "antenna"

Most systems also have a jack for connecting headphones, and some have the control box below the shaft and a small display unit above.
Operating a metal detector is simple. Once you turn the unit on, you move slowly over the area you wish to search. In most cases, you sweep the coil (search head) back and forth over the ground in front of you. When you pass it over a target object, an audible signal occurs. More advanced metal detectors provide displays that pinpoint the type of metal it has detected and how deep in the ground the target object is located.
The circuit diagram of a typical metal detector is shown in **figure 2**.

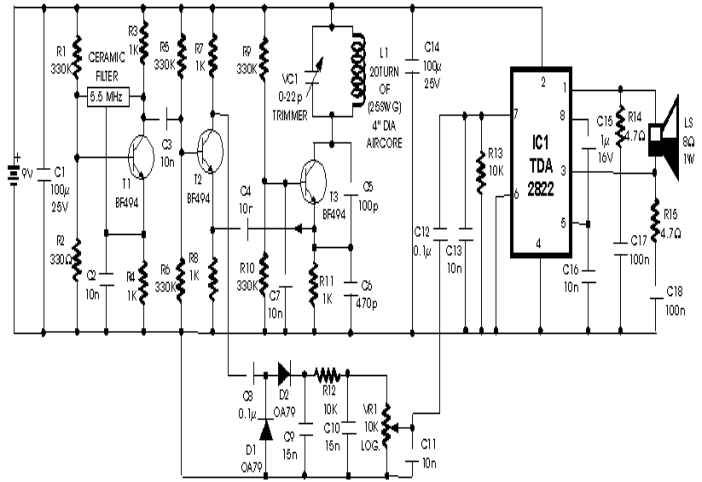

**Figure 2:** *The Circuit Diagram of a Metal Detector with Alarm (source: www.electrokits.com)*

The metal detector shown in **Figure 2** is a metal detector with a 9V DC source and a 1W alarm system.

**3.2.0 Pressure Sensitive mat (The Intrusion Detection Layer):**

The pressure mat incorporated in this paper is the SMS 3 safety mat. It is designed to safeguard personnel when entering a hazardous area around dangerous machinery.
Individual's presence is detected upon their walking on the mat; the interconnected safety controller continuously monitors the integrity of the safety mat system, sending a stop signal in the event of a system fault or pressure mat actuation. (www.schmersalusa.com)

**3.2.1. ANATOMY OF THE SMS 3 SAFETY PRESSURE MAT**

According to www.schmersalusa.com, a typical series SMS 3 safety pressure mat comprises of the following:

- Non-slip safety mat surface
- Upper electrode (24-guage steel plate, hardened for optimum performance & durability)
- Edge spacer

- Conductive (u-shaped) contact strips
- Compressible, elastomeric insulting strips
- Lower electrode (24-guage steel plate, hardened for optimum performance &durability)
- Safety mat bottom surface

A few added advantages of the SMS 3 safety pressure mat been chosen are:

- It is easy-to-install simple 4-wire connection (with no need of terminal resistor or additional base plate).
- Its 6m cable satisfies wiring requirements for a wide variety of applications.
- Its non-slip surface enhances traction and minimize slippage alert.

### 3.3.0.   Electronic Access Control (The Mechanical/ Electronic Access Control Layer)

A microcontroller chip PIC16F873 controls the activities of the access control system. A schematic circuit diagram is shown in figure 3.

The Power Section: it consists of a step down transformer used to step the voltage down to 5V which is the required voltage to run the Digital Access System. The Key Pads are used to enter the PIN number. It also has Reset, Open, Change Pass, Cancel and Default buttons. It also has number buttons. It has an administrative password and a user password. The administrative pass can be used to reset the password. The default password is the administrative password and can only be accessed from inside the room. The microcontroller has two ROMs of which one is used to store the PIN and the other is used to compare the stored PIN with the entered PIN. The relay section is connected to the mag lock.

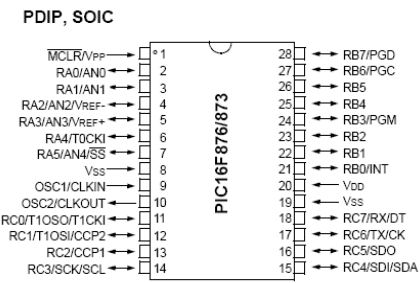

**Figure 4:** *The PIC16F873 Microcontroller Pin Functions (source: PIC16F87x Data Sheet)*

**Table 1:** *The Key Features of the PIC16F873 Microcontroller (source: PIC16F87x Data Sheet)*

| Key Features PICmicro™ Mid-Range Reference Manual (DS33023) | PIC16F873 |
|---|---|
| Operating Frequency | DC - 20 MHz |
| RESETS (and Delays) | POR, BOR (PWRT, OST) |
| FLASH Program Memory (14-bit words) | 4K |
| Data Memory (bytes) | 192 |
| EEPROM Data Memory | 128 |
| Interrupts | 13 |
| I/O Ports | Ports A,B,C |
| Timers | 3 |
| Capture/Compare/PWM Modules | 2 |
| Serial Communications | MSSP, USART |
| Parallel Communications | — |
| 10-bit Analog-to-Digital Module | 5 input channels |
| Instruction Set | 35 instructions |

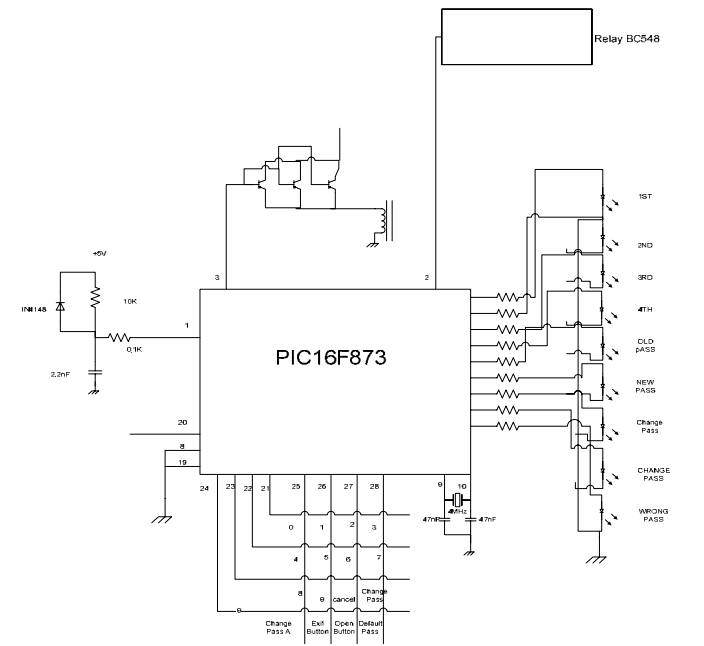

**Figure 3:** *A Schematic Circuit Diagram showing the implementation of the Microcontroller PIC16F873 in a Digital Access System*

From **Table 1**, the PIC16F873 microcontroller several features which are briefly explained as follows:

**Memory:** Data, FLASH Program and EEPROM data memories are the three types of memory found in the microcontroller. Data memory is capable of storing 192 bytes of data. FLASH Program memory stores 4kb of 14-bit words. EEPROM is a form of memory which can be written upon, read from and erased after use. This makes it a useful form of memory form continuous or flexible usage.

**I/O Ports:** these are ports used for inputting and outputting information to and from the microcontroller. These are used to supply the information to be processed and the results to the appropriate devices.

**Serial and Parallel Communications:** these are used for passing information between devices and the microcontroller.

**Timers:** used for synchronizing and ordering the sequence of events.

**Instruction Set:** microcontrollers are RISC (Reduced Instruction Set Codes) devices. They have reduced number of Instruction sets and are considerably fast.

**Interrupts, RESET and Delays:** these are signal codes used to notify users of completed tasks, restart the system or delay the operations of the system due to ongoing activities.

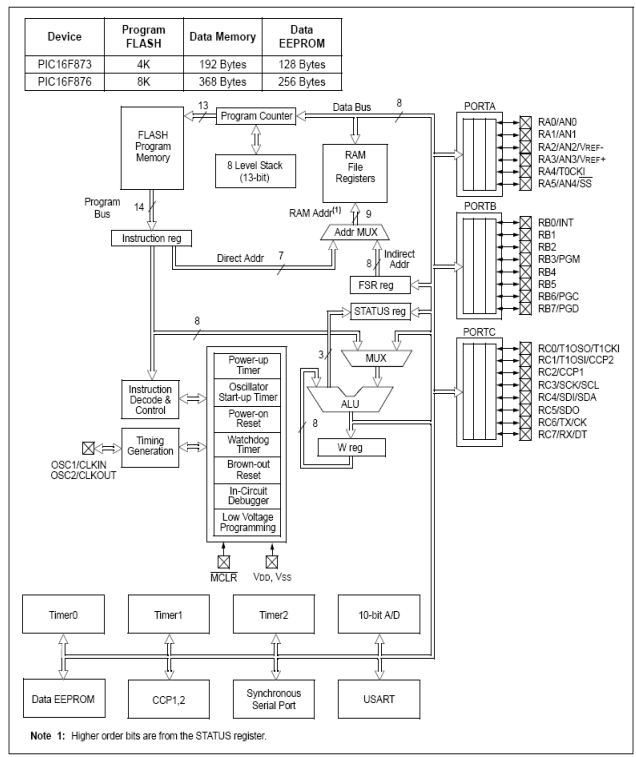

**Figure 5:** *The Block Diagram of the PIC16F873 Microcontroller (source: PIC16F87x Data Sheet)*

**Table 2:** *Table Showing the Pin Functions of the PIC16F873 Microcontroller (source: PIC16F87x Data Sheet)*

| Pin Name | DIP Pin# | SOIC Pin# | I/O/P Type | Buffer Type | Description |
|---|---|---|---|---|---|
| OSC1/CLKIN | 9 | 9 | I | ST/CMOS[3] | Oscillator crystal input/external clock source input. |
| OSC2/CLKOUT | 10 | 10 | O | — | Oscillator crystal output. Connects to crystal or resonator in crystal oscillator mode. In RC mode, the OSC2 pin outputs CLKOUT which has 1/4 the frequency of OSC1, and denotes the instruction cycle rate. |
| MCLR/Vpp | 1 | 1 | I/P | ST | Master Clear (Reset) input or programming voltage input. This pin is an active low RESET to the device. |
| | | | | | PORTA is a bi-directional I/O port. |
| RA0/AN0 | 2 | 2 | I/O | TTL | RA0 can also be analog input0. |
| RA1/AN1 | 3 | 3 | I/O | TTL | RA1 can also be analog input1. |
| RA2/AN2/VREF- | 4 | 4 | I/O | TTL | RA2 can also be analog input2 or negative analog reference voltage. |
| RA3/AN3/VREF+ | 5 | 5 | I/O | TTL | RA3 can also be analog input3 or positive analog reference voltage. |
| RA4/T0CKI | 6 | 6 | I/O | ST | RA4 can also be the clock input to the Timer0 module. Output is open drain type. |
| RA5/SS/AN4 | 7 | 7 | I/O | TTL | RA5 can also be analog input4 or the slave select for the synchronous serial port. |
| | | | | | PORTB is a bi-directional I/O port. PORTB can be software programmed for internal weak pull-up on all inputs. |
| RB0/INT | 21 | 21 | I/O | TTL/ST[1] | RB0 can also be the external interrupt pin. |
| RB1 | 22 | 22 | I/O | TTL | |
| RB2 | 23 | 23 | I/O | TTL | |
| RB3/PGM | 24 | 24 | I/O | TTL | RB3 can also be the low voltage programming input. |
| RB4 | 25 | 25 | I/O | TTL | Interrupt-on-change pin. |
| RB5 | 26 | 26 | I/O | TTL | Interrupt-on-change pin. |
| RB6/PGC | 27 | 27 | I/O | TTL/ST[2] | Interrupt-on-change pin or In-Circuit Debugger pin. Serial programming clock. |
| RB7/PGD | 28 | 28 | I/O | TTL/ST[2] | Interrupt-on-change pin or In-Circuit Debugger pin. Serial programming data. |
| | | | | | PORTC is a bi-directional I/O port. |
| RC0/T1OSO/T1CKI | 11 | 11 | I/O | ST | RC0 can also be the Timer1 oscillator output or Timer1 clock input. |
| RC1/T1OSI/CCP2 | 12 | 12 | I/O | ST | RC1 can also be the Timer1 oscillator input or Capture2 input/Compare2 output/PWM2 output. |
| RC2/CCP1 | 13 | 13 | I/O | ST | RC2 can also be the Capture1 input/Compare1 output/PWM1 output. |
| RC3/SCK/SCL | 14 | 14 | I/O | ST | RC3 can also be the synchronous serial clock input/output for both SPI and I2C modes. |
| RC4/SDI/SDA | 15 | 15 | I/O | ST | RC4 can also be the SPI Data In (SPI mode) or data I/O (I2C mode). |
| RC5/SDO | 16 | 16 | I/O | ST | RC5 can also be the SPI Data Out (SPI mode). |
| RC6/TX/CK | 17 | 17 | I/O | ST | RC6 can also be the USART Asynchronous Transmit or Synchronous Clock. |
| RC7/RX/DT | 18 | 18 | I/O | ST | RC7 can also be the USART Asynchronous Receive or Synchronous Data. |
| Vss | 8, 19 | 8, 19 | P | — | Ground reference for logic and I/O pins. |
| Vdd | 20 | 20 | P | — | Positive supply for logic and I/O pins. |

Legend: I = input   O = output   I/O = input/output   P = power
— = Not used   TTL = TTL input   ST = Schmitt Trigger input
Note 1: This buffer is a Schmitt Trigger input when configured as the external interrupt.
2: This buffer is a Schmitt Trigger input when used in Serial Programming mode.
3: This buffer is a Schmitt Trigger input when configured in RC oscillator mode and a CMOS input otherwise.

### 3.4. MODE OF OPERATION OF THE PROPOSED SYSTEM.

The metal detector representing the first phase of security (Environmental Layer), would be an independent unit which could either be a hand-held or a standing unit metal detector with security guards monitoring access to the area. The security guards also serve as a part of the environmental layer. Once an individual is cleared through the first layer of security, he is permitted into an infusion of the second and third phase (the integration of the pressure mat and the electronic access control), as both would work asynchronously. This implies that once access is denied to the individual by the electronic access control, the pressure mat would be activated and on sensing of pressure, the alarm is activated. An OR gate IC will be used to link the outputs from the pressure sensitive mat and the output from the Electronic Access System to the alarm. The metal detector is not included due to the fact that the metal detector has an alarm of its own. The system also has an alarm system. The alarm system could be a blow horn or a continuously blinking light connected to the system. A blow horn can be implemented in a situation where the intruder needs to be scared off. The continuously blinking light can be an alert set at the security post in a situation where the intruder needs to be caught and interrogated. In conclusion, an individual has to pass the metal detector phase, and then pass the integrated electronic access control and pressure sensitive mat phase for access to be granted.

## 4. LIMITATIONS

This proposed design is however limited in the following ways:

- A fault in one of the system continuously gives false alarms which can be frustrating for the monitoring officer.
- A surveillance system is system which can only be used as a reference after an intrusion has occurred. This can be bypassed by a well clothed thief.
- Monitoring surveillance tapes can also be very difficult. This is because the monitoring officer cannot at all times catch everything going on. This will require many monitoring officers.
- Authentication systems can be bypassed by digitally cracking the codes.
- This system is a security system and lacks safety features so avert danger from within.

### 4.1. Possible Solutions to the Limitations of the System

- False alarms and irresponsive alarms can be as a result of bouncing signals. Bouncing signals are common to equipment using push down buttons, such as this system, in which continuous pressing of the buttons results in distortions of the pulse signals. This can either trigger the alarm or not. This can however be reduced or stopped by using a de-bouncing circuit. A de-bouncing circuit is a connection of a buffer, a resistor and a capacitor. This circuit ensures the generation of a regular pulse.
- Cracking the codes for an authentication system will require time. Some hackers try hacking by "trial by error" while others use devices capable of cracking codes within a short time. Whatever method used still requires time and contact with the device. The use of security personnel will tend to dissuade such attacks.
- The system can be infused will safety features such as smoke detectors and other safety features to reduce the risk of having internal damage.

## 5. CONCLUSION

In this paper we have presented a security system designed to help in the safe-guard of important materials and also to alert the appropriate personnel's about unauthorized access to the room. This paper outlines the important factors to be considered in developing a security system. This paper discusses design of a security system using already existing security devices. This system is however limited by certain factors and the possible solutions to these limitations are also discussed.

**Biographies**

**Awodele Oludele** is a presently a lecturer in the Computer Science and Mathematics Department, Babcock University, Ilishan-Remo, Ogun State, Nigeria. He recently completed his Ph.D programme in Computer Science in the University of Agriculture Abeokuta, Ogun state, Nigeria. His research areas are Software Engineering, Data Communication and Artificial Intelligence. He has published works in several journals of international repute.

**Omole Oladele** is a final year undergraduate student of Computer Technology at Babcock University, Ilishan-Remo, Ogun state, Nigeria. He is a member of the Young Engineer's Team. His areas of interest include Networking and Gaming. He is a Microsoft Certified Professional.

**Ogunnusi Ayodele** is a final year undergraduate student of Computer Technology at Babcock University, Ilishan-Remo, Ogun state, Nigeria. He holds a Microsoft Certified Professional certificate. His area of interest includes web design, networking and project management.

**Seton Olurotimi** is a final year undergraduate student of Computer Technology at Babcock University, Ilishan-Remo, Ogun state, Nigeria. He is a member of Young Engineer's Team. His areas of interest include Electrical Circuit Design and physics. He is a Microsoft Certified Professional certificate holder.